\documentclass[12pt]{article}

\usepackage{amsmath,amssymb, amsthm}

\title{Extended Tensor Products and Generalization of the Notion of Entanglement}

\author{Andrei Khrennikov\\
International Center for Mathematical Modelling
\\in Physics and Cognitive Sciences\\
Linnaeus University,  V\"axj\"o, Sweden\\
Elemer E. Rosinger\\
  Department of Mathematics, University of Pretoria\\
South-African Republic}

\begin{document}

\maketitle

\begin{abstract}
Motivated by the novel applications of the mathematical formalism of quantum theory and its generalizations in cognitive science, 
psychology, social and political sciences, and economics, we extend the notion of the tensor product and entanglement. We also 
study the relation between conventional entanglement of complex qubits and our generalized entanglement.
Our construction can also be used to describe entanglement in the framework of non-Archimedean physics. It is also possible to construct tensor products of non-Archimedean (e.g., $p$-adic) and complex Hilbert spaces.
\end{abstract}

\section{Introduction}

The notion of the tensor product plays a crucial role in quantum theory: states of composite quantum systems belong to the tensor
product of Hilbert state spaces of subsystems. The tensor product is the basic structure of quantum information theory; in particular, such a fundamental notion as entanglement is defined on the basis of the tensor product. Recently the quantum formalism and
its generalization started to be applied outside of the domain of quantum physics: in cognitive science, psychology, economics,
and finances, see, e.g., \cite{KC}, \cite{A} and bibliography in these works. In these applications it became clear that the conventional quantum formalism is too special and restricted to cover all these novel applications. More general quantum-like models describing 
nonclassical flows of information are on demand. In particular, composite cognitive and social systems need not be described
by the conventional tensor product; they may exhibit other types of entanglement which are not reduced to the conventional quantum entanglement. This was one of motivations for E. Rosinger to extend the notion of the conventional tensor product \cite{R} and entanglement. In this note we continue the activity in this direction. The tensor construction of \cite{R} is extened even further
to obtain a possibility to use operator algebras, instead of fields of scalars. We consider a few examples related to generalized
tensor representation of the standard Bell states. Among other possible applications of the extended tensor product we can mention
the description of generalized quantum systems which state spaces are not simply complex Hilbert spaces, but (in general noncommutative)
groups. We can also describe entanglement in the framework of non-Archimedean physics, e.g.,  \cite{NA},  \cite{R1}. It is also possible to construct 
tensor products of non-Archimedean (e.g., $p$-adic) and complex Hilbert spaces. The latter is a subject of our further investigations.

\section{Definition of extended tensor products}

Let \\

$\alpha : X \times X \longrightarrow X,~ \beta : Y \times Y \longrightarrow Y$ \\

be two arbitrary maps.

Let us denote by $Z$ the set of all finite sequences of pairs \\

(1)~~~ $ ( x_1, y_1 ), \dots , ( x_n, y_n ) $ \\

where $n \in {\bf N}_1$, while $x_i \in X,~ y_i \in Y$, with $1 \leq i \leq n$. \\

We define on $Z$ the binary operation $\gamma$ simply by the concatenation of the sequences (1). It follows that $\gamma$ is associative, therefore, each sequence (4) can be written as \\

(2)~~~ $ ( x_1, y_1 ), \dots , ( x_n, y_n ) =
                 ( x_1, y_1 ) \gamma ( x_2, y_2 ) \gamma \ldots \gamma ( x_n, y_n ) $ \\

where for $n = 1$, the right hand term is understood to be simply $( x_1, y_1 )$. Obviously, if $X$ or $Y$ have at least two elements, then $\gamma$ is not commutative. \\

Then (1), (2) give \\

(3)~~~ $ Z = \left \{ ( x_1, y_1 ) ~\gamma~ ( x_2, y_2 ) ~\gamma~ \ldots ~\gamma~ ( x_n, y_n ) ~~
                             \begin{array}{|l}
                              ~ n \geq 1 \\ \\
                              ~ x_i \in X,~ y_i \in Y,~ 1 \leq i \leq n
                             \end{array} \right \} $ \\ \\

which obviously results in \\

(4)~~~ $ X \times Y \subseteq Z $ \\

Now we define on $Z$ an equivalence relation $\approx_{\alpha, \beta}$ as follows. Two sequences in (1) are equivalent, if and only if they are identical, or each can be obtained from the other by a finite number of applications of the following operations : \\

(5) permuting the pairs $( x_i, y_i )$ within the sequence \\

(6) replacing $( \alpha ( x_1, x\,'_1 ) , y_1 ), ( x_2, y_2 ), \ldots , ( x_n, y_n )$ with \\
      \hspace*{0.8cm} $( x_1, y_1 ), ( x\,'_1, y_1 ), ( x_2, y_2 ), \ldots , ( x_n, y_n )$, or
      vice-versa \\

(7) replacing $( x_1, \beta ( y_1, y\,'_1 ) ), ( x_2, y_2 ), \ldots ,( x_n, y_n )$ with \\
       \hspace*{1cm} $( x_1, y_1 ), ( x_1, y\,'_1 ), ( x_2, y_2 ), \ldots , ( x_n, y_n )$, or
       vice-versa \\

Let us note that, in view of the rather general related result in Lemma 1. below, the binary relation $\approx_{\alpha, \beta}$ defined above on $Z$ is indeed an equivalence relation. \\

Finally, the {\it tensor product} of $( X, \alpha )$ and $( Y, \beta )$ is defined to be the quotient space \\

(8)~~~ $ X \bigotimes_{\alpha, \beta} Y = Z / \approx_{\alpha, \beta} $ \\

with the mapping $\tau_{\alpha, \beta}$ induced through the inclusion (4) by the canonical quotient embedding
corresponding to (8), namely \\

(9)~~~ $ X \times Y \ni ( x, y ) \stackrel{\tau_{\alpha, \beta}}\longmapsto
                           x \bigotimes_{\alpha, \beta} y \in X \bigotimes_{\alpha, \beta} Y $ \\

where as in the usual case of tensor products, we denote by $x \bigotimes_{\alpha, \beta} y$, or simply $x \bigotimes y$, the $\approx_{\alpha, \beta}$ equivalence class of $( x, y )$. \\

Furthermore, the equivalence $\approx_{\alpha, \beta}$ is {\it compatible} with the semigroup structure $( Z, \gamma )$, thus (8) has in fact the stronger form which gives a {\it commutative semigroup} structure on the resulting generalized tensor product $X \bigotimes_{\alpha, \beta} Y$, namely \\

(10)~~~ $ ( X \bigotimes_{\alpha, \beta} Y, \gamma / \approx_{\alpha, \beta} ) =
                                                 ( Z, \gamma ) / \approx_{\alpha, \beta} $ \\

For simplicity, however, we shall write $\gamma$ instead of $\gamma / \approx_{\alpha, \beta}$. \\

In this way, the elements of $X \bigotimes_{\alpha, \beta} Y$ are all the expressions \\

(11)~~~ $ x_1 \bigotimes_{\alpha, \beta} y_1 ~\gamma~ x_2 \bigotimes_{\alpha, \beta} y_2 ~\gamma~
                                                   \ldots ~\gamma~ x_n \bigotimes_{\alpha, \beta} y_n $ \\

with $n \geq 1$ and $x_i \in X,~ y_i \in Y$, for $1 \leq i \leq n$. \\ \\

\section{How large are the extended tensor products ?} 

Before going further, let us see when is the mapping (9) {\it injective}. A {\it necessary} condition is given by \\

{\bf Proposition 1.} \\

If the mapping $\tau_{\alpha, \beta}$ in (9), namely \\

(12)~~~ $ X \times Y \ni ( x, y ) \stackrel{\tau_{\alpha, \beta}}\longmapsto x
                                    \bigotimes_{\alpha, \beta} y \in X \bigotimes_{\alpha, \beta} Y $ \\

is {\it injective}, then the binary operations $\alpha$ and $\beta$ are {\it associative}. \\

{\bf Proof.} \\

We first show that \\

(13)~~~ $ \alpha $~ not associative $~ \Longrightarrow~ \tau_{\alpha, \beta} $~ not injective \\

Indeed, let $a, b, c \in X$, such that $d = \alpha ( \alpha ( a, b ), c ) \neq \alpha ( a, \alpha ( b, c ) ) = e$. Further,
let $y \in X$. Then in view of (6), we have \\

(14)~~~$ \begin{array}{l}
                 ( d, y ) = ( \alpha ( \alpha ( a, b ) ), c ), y )
                 \approx_{\alpha, \beta} ~( \alpha ( a, b ), y ) ~\gamma~ ( c, y ) \approx_{\alpha, \beta} \\ \\

                 ~~~~~~ \approx_{\alpha, \beta} ( a, y ) ~\gamma~ ( b, y ) ~\gamma~ ( c, y) \\ \\
                 ~~~~~~ \approx_{\alpha, \beta} ( a, y ) ~\gamma~ ( \alpha ( b, c ), y ) \approx_{\alpha, \beta} \\ \\

                 ~~~~~~ \approx_{\alpha, \beta} ( \alpha ( a, \alpha ( b, c )), y ) = ( e, y )
           \end{array} $ \\

hence $( d, y ) \approx_{\alpha, \beta} ( e, y )$, while obviously $( d, y ) \neq ( e, y )$. \\

In a similar manner, we also have \\

(15)~~~ $ \beta $~ not associative $~ \Longrightarrow~ \tau_{\alpha, \beta} $~ not injective \\

The converse of Proposition 3.1. does {\it not} hold, as illustrated in \\

{\bf Example 1.} \\

The above definition contains as a particular case the usual tensor products of groups. And for Abelian groups one has \\

$~~~~~~ {\bf Z}/ ( m ) \bigotimes_{\bf Z}{\bf Z}/ ( n ) = {\bf Z}/ ( d ) $ \\

for $m, n \in {\bf N}$, and $d$ the greatest common divisor of $m$ and $n$. Thus in particular \\

$~~~~~~ {\bf Z}/ ( 2 ) \bigotimes_{\bf Z}{\bf Z}/ ( 3 ) = 0 $

Clearly, the binary operation $\gamma$ on $Z$ will canonically lead by this quotient operation to a {\it commutative} and {\it associative} binary operation on $X \bigotimes_{\alpha, \beta} Y$, which for convenience is denoted by the same $\gamma$, although this time it depends on $\alpha$ and $\beta$. \\

The customary and highly particular situation is when $X$ and $Y$ are semigroups, groups, or even vector spaces over some field ${\bf K}$. In this case $\alpha, \beta$ and $\gamma$ are as usual denoted by +, that is, the sign of addition. \\

It is easy to note that in the construction of tensor products above, it is {\it not} necessary for $( X, \alpha )$ and $( Y, \beta )$ to be semigroups, let alone groups, or for that matter, vector spaces. Indeed, it is sufficient that $\alpha$ and $\beta$ are arbitrary binary operations on $X$ and $Y$, respectively. \\

Also, as seen above, $\alpha$ and $\beta$ need {\it not} be commutative either. However, the tensor product $X
\bigotimes_{\alpha, \beta} Y$, with the respective binary operation $\gamma$, will nevertheless be commutative and
associative. \\

It is important to note that the tensor products defined above have a {\it universality} property which is a natural generalization of the corresponding well known one for usual tensor products. \\ \\

\section{Extended concepts of entanglement}

{\bf Definition 1.} \\

Given two binary operations $\alpha : X \times X \longrightarrow X$ and $\beta : Y \times Y \longrightarrow Y$. An element $w \in X \bigotimes_{\alpha, \beta} Y$ is called {\it entangled}, if and only if it is {\it not} of the form \\

(16)~~~ $ w = x \bigotimes_{\alpha, \beta} y $ \\

for some $x \in X$ and $y \in Y$. \\

{\bf Note 1.} \\

1) Since it was noted that the usual tensor products are particular cases of the tensor products defined in this section, it follows that the definition of entanglement given above does indeed generalize the usual concept of entanglement. \\

2) It is important to note that generalized tensor products (8) can have an interest even when the corresponding
mappings (9) are not injective. Indeed, if for instance in such cases one still has the {\it strict} inclusion \\

(17)~~~ $ \tau_{\alpha, \beta} ( X \times Y ) \subset X \bigotimes_{\alpha, \beta} Y, \tau_{\alpha, \beta} ( X \times Y ) \not= X \bigotimes_{\alpha, \beta} Y $ \\

then there are still {\it entangled} elements in $X \bigotimes_{\alpha, \beta} Y$, namely, those in the nonvoid set \\

(18)~~~ $ X \bigotimes_{\alpha, \beta} Y ~\setminus~ \tau_{\alpha, \beta} ( X \times Y ) $ \\

3) As seen in elsewhere, tensor products can be defined in far more general ways than above. And with such far more general definitions there are plenty of cases when the mappings corresponding to (9) will be injective.

In the construction of tensor products above, we used the following easy to prove \\

{\bf Lemma 1.} \\

Let on a nonvoid set $E$ be given a family $( \equiv_i )_{i \in I}$ of {\it symmetric} binary relations. Further, let us define on $E$ the binary relation $\approx$ as follows. For $a, b \in E$, we have $a \approx b$, if and only if $a = b$, or there exists a finite sequence \\

$~~~~~~ a = c_0 \equiv_{i_0} c_1 \equiv_{i_1} c_2 \equiv_{i_2} \ldots \equiv_{i_{n-2}} c_{n-1} \equiv_{i_{n-1}} c_n = y $ \\

where $c_1, \ldots , c_{n-1} \in E$. \\

Then $\approx$ is an {\it equivalence} relation on $E$. \\ \\

\section{Further ways to extend the concepts of tensor products and entanglements}

Let \\

${\cal A} \subseteq X^X,~ {\cal B} \subseteq Y^Y$ \\

be arbitrary mappings. Then we define \\

${\cal C} \subseteq Z^Z$ \\

as the set of all pairs of mappings \\

$C = ( A, B ) \in {\cal A} \times {\cal B}$ \\

which act according to \\

$Z \ni ( x_1, y_1 ), \dots , ( x_n, y_n ) ~~\longmapsto~~ \\ \\
               ~~~~~~~~~~~~~~~~~~~~~\longmapsto~~ ( A ( x_1 ), B ( y_1 ) ), \dots , ( A ( x_n ), B ( y_n ) ) \in Z$ \\

Now we consider the particular case when \\

$X = Y,~ {\cal A} = {\cal B}$ \\

and then we define on $Z$ the equivalence relation $\approx_{\cal A}$ as follows. Two sequences in (1) are equivalent, if and only if they are identical, or each can be obtained from the other by a finite number of applications of the following operations : \\

(19) permuting the pairs $( x_i, y_i )$ within the sequence \\

(20) replacing $( A ( x_1 ) , B ( y_1 ) ), \ldots , ( A ( x_n ), B ( y_n ) )$ with \\
      \hspace*{0.8cm} $( B A ( x_1 ), y_1 ), \ldots , ( B A ( x_n ), y_n )$, or with \\
      \hspace*{0.8cm} $( x_1 , A B ( y_1 ) ), \ldots , ( x_n, A B ( y_n ) )$, or with \\
      \hspace*{0.8cm} $( A, B ) ( ( x_1, y_1 ), \ldots , ( x_n, y_n ) )$, or vice-versa, \\
      \hspace*{0.8cm} where $A, B \in {\cal A}$ \\

In view of the above Lemma 1., it follows that $\approx_{\cal A}$ is indeed an equivalence relation on $Z$, therefore, we can define the tensor product \\

(21)~~~~ $ X \bigotimes_{\cal A} X = Z / \approx_{\cal A} $ \\

with the mapping $\tau_{\cal A}$ induced through the inclusion (4) by the canonical quotient embedding
corresponding to (21), namely \\

(22)~~~ $ X \times X \ni ( x, y ) \stackrel{\tau_{\cal A}} \longmapsto
                           x \bigotimes_{\cal A} y \in X \bigotimes_{\cal A} Y $ \\

where as in the usual case of tensor products, we denote by $x \bigotimes_{\cal A} y$, or simply $x \bigotimes y$, the $\approx_{\cal A}$ equivalence class of $( x, y )$. \\

Furthermore, the equivalence $\approx_{\cal A}$ is {\it compatible} with the semigroup structure $( Z, \gamma )$, thus (21) has in fact the stronger form which gives a {\it commutative semigroup} structure on the resulting generalized tensor product $X \bigotimes_{\cal A} X$, namely \\

(23)~~~ $ ( X \bigotimes_{\cal A} X, \gamma / \approx_{\cal A} ) = ( Z, \gamma ) / \approx_{\cal A} $ \\

For simplicity, however, we shall write $\gamma$ instead of $\gamma / \approx_{\cal A}$. \\

In this way, the elements of $X \bigotimes_{\cal A} X$ are all the expressions \\

(24)~~~ $ x_1 \bigotimes_{\cal A} y_1 ~\gamma~ x_2 \bigotimes_{\cal A} y_2 ~\gamma~
                                                   \ldots ~\gamma~ x_n \bigotimes_{\cal A} y_n $ \\

with $n \geq 1$ and $x_i, y_i \in X$, for $1 \leq i \leq n$. \\ \\

\section{Further extensions of the concept of entanglement} 

The above tensor products \\

$X \bigotimes_{\cal A} X$ {\bf and} $X \bigotimes_{\alpha, {\cal A}} X$ \\

have corresponding concepts of entanglement according to obvious extensions of Definition 1. above. \\ \\

{\bf A Mixture of the Above} \\

Let \\

$\alpha : X \times X \longrightarrow X$ \\

and \\

${\cal A} \subseteq X^X$ \\

and consider \\

${\cal C} \subseteq Z^Z$ \\

as the set of all pairs of mappings \\

$C = ( A, B ) \in {\cal A} \times {\cal A}$ \\

which act according to \\

$Z \ni ( x_1, y_1 ), \dots , ( x_n, y_n ) ~~\longmapsto~~ \\ \\
          ~~~~~~~~~~~~~~~~~\longmapsto~~ ( A ( x_1 ), B ( y_1 ) ), \dots , ( A ( x_n ), B ( y_n ) ) \in Z$ \\

Define now on $Z$ the equivalence relation $\approx_{\alpha, {\cal A}}$ as follows. Two sequences in (1) are equivalent, if and only if they are identical, or each can be obtained from the other by a finite number of applications of the following operations : \\

(25) permuting the pairs $( x_i, y_i )$ within the sequence \\

(26) replacing $( \alpha ( x_1, x\,'_1 ) , y_1 ), ( x_2, y_2 ), \ldots , ( x_n, y_n )$ with \\
      \hspace*{0.8cm} $( x_1, y_1 ), ( x\,'_1, y_1 ), ( x_2, y_2 ), \ldots , ( x_n, y_n )$, or
      vice-versa \\

(27) replacing $( x_1, \alpha ( y_1, y\,'_1 ) ), ( x_2, y_2 ), \ldots ,( x_n, y_n )$ with \\
       \hspace*{1cm} $( x_1, y_1 ), ( x_1, y\,'_1 ), ( x_2, y_2 ), \ldots , ( x_n, y_n )$, or
       vice-versa \\

(28) replacing $( A ( x_1 ) , B ( y_1 ) ), \ldots , ( A ( x_n ), B ( y_n ) )$ with \\
      \hspace*{0.8cm} $( B A ( x_1 ), y_1 ), \ldots , ( B A ( x_n ), y_n )$, or with \\
      \hspace*{0.8cm} $( x_1 , A B ( y_1 ) ), \ldots , ( x_n, A B ( y_n ) )$, or with \\
      \hspace*{0.8cm} $( A, B ) ( ( x_1, y_1 ), \ldots , ( x_n, y_n ) )$, or vice-versa, \\
      \hspace*{0.8cm} where $A, B \in {\cal A}$ \\

Then in view of Lemma 1. above, it follows that $\approx_{\alpha, {\cal A}}$ is indeed an equivalence relation on $Z$, therefore, we can define the tensor product \\

(29)~~~~ $ X \bigotimes_{\alpha, {\cal A}} X = Z / \approx_{\alpha, {\cal A}} $ \\

with the mapping $\tau_{\cal A}$ induced through the inclusion (4) by the canonical quotient embedding
corresponding to (31), namely \\

(30)~~~ $ X \times X \ni ( x, y ) \stackrel{\tau_{\alpha, {\cal A}}} \longmapsto
                           x \bigotimes_{\alpha, {\cal A}} y \in X \bigotimes_{\alpha, {\cal A}} Y $ \\

where as in the usual case of tensor products, we denote by $x \bigotimes_{\alpha, {\cal A}} y$, or simply $x \bigotimes y$, the $\approx_{\alpha, {\cal A}}$ equivalence class of $( x, y )$. \\

Furthermore, the equivalence $\approx_{\alpha, {\cal A}}$ is {\it compatible} with the semigroup structure $( Z, \gamma )$, thus (29) has in fact the stronger form which gives a {\it commutative semigroup} structure on the resulting generalized tensor product $X \bigotimes_{\alpha, {\cal A}} X$, namely \\

(31)~~~ $ ( X \bigotimes_{\alpha, {\cal A}} X, \gamma / \approx_{\alpha, {\cal A}} ) = ( Z, \gamma ) /\approx_{\alpha,
                       {\cal A}} $ \\

For simplicity, however, we shall write $\gamma$ instead of $\gamma / \approx_{\alpha, {\cal A}}$. \\

In this way, the elements of $X \bigotimes_{\alpha, {\cal A}} X$ are all the expressions \\

(32)~~~ $ x_1 \bigotimes_{\alpha, {\cal A}} y_1 ~\gamma~ x_2 \bigotimes_{\alpha, {\cal A}} y_2 ~\gamma~
                                                   \ldots ~\gamma~ x_n \bigotimes_{\alpha, {\cal A}} y_n $ \\

with $n \geq 1$ and $x_i, y_i \in X$, for $1 \leq i \leq n$. \\ \\

\section{Relationship between the three tensor products} 

$X \bigotimes_{\alpha, \alpha} X$,~ $X \bigotimes_{\cal A} X$ {\bf and} $X \bigotimes_{\alpha, {\cal A}} X$ \\

In view of Lemma 2. below, we have the {\it surjective} mapping, see (11) \\

(33)~~~ $ \lambda_{\alpha, {\cal A}} : X \bigotimes _{\alpha, \alpha} X \ni x_1 \bigotimes_{\alpha, \alpha} y_1 ~\gamma~ x_2 \bigotimes_{\alpha, \alpha} y_2 ~\gamma~ \ldots ~\gamma~ x_n \bigotimes_{\alpha, \alpha} y_n ~~\longmapsto~~ \\ \\
~~~~~~~~~~~~~~\longmapsto~~ x_1 \bigotimes_{\alpha, {\cal A}} y_1 ~\gamma~ x_2 \bigotimes_{\alpha, {\cal A}} y_2 ~\gamma~ \ldots ~\gamma~ x_n \bigotimes_{\alpha, {\cal A}} y_n \in X \bigotimes _{\alpha, {\cal A}} X $ \\

as well as the {\it surjective} mapping, see (24) \\

(34)~~~ $ \mu_{\alpha, {\cal A}} : X \bigotimes _{\cal A} X \ni x_1 \bigotimes_{\cal A} y_1 ~\gamma~ x_2 \bigotimes_{\cal A} y_2 ~\gamma~ \ldots ~\gamma~ x_n \bigotimes_{\cal A} y_n ~~\longmapsto~~ \\ \\
~~~~~~~~~~~~~~\longmapsto~~ x_1 \bigotimes_{\alpha, {\cal A}} y_1 ~\gamma~ x_2 \bigotimes_{\alpha, {\cal A}} y_2 ~\gamma~ \ldots ~\gamma~ x_n \bigotimes_{\alpha, {\cal A}} y_n \in X \bigotimes _{\alpha, {\cal A}} X $ \\ \\

{\bf A Further Lemma} \\

{\bf Lemma 2.} \\

Let on a nonvoid set $E$ be given two families $( \equiv_i )_{i \in I}$ and $( \equiv_j )_{j \in J}$ of {\it symmetric} binary relations, where $I \subseteq J$. Let us denote by $\approx_I$ and $\approx_j$ the respective equivalence relations in $E$ constructed according to Lemma 1. Then for $a, b \in E$, we have \\

(35)~~~ $ a \approx_I b ~~\Longrightarrow~~ a \approx_J b $ \\

therefore, we have the {\it surjective} mapping \\

(36)~~~ $ E / \approx_I \, \, \ni ( a )_{\approx_I} ~\longmapsto~ ( a )_{\approx_J} \in E / \approx_J $ \\

where $( a )_{\approx_I}$ and $( a )_{\approx_J}$ are the equivalence classes of $a \in E$ with respect to the corresponding equivalence relations $\approx_I$ and $\approx_J$. \\

{\bf Note 2.} \\

The above relation (20) means that by {\it enlarging} the family $( \equiv_i )_{i \in I}$ of symmetric binary relations on a set $E$, one {\it decreases} the corresponding quotient space $E / \approx_I $. \\ \\

{\bf Examples of generalized tensor product representations of the conventional Bell's states:} 

Let  $X = Y = H = {\bf C}^2, \alpha = \beta = +$ on $H,$
${\cal A} = {\cal B} = L ( H )$ the algebra of linear operators on the Hilbert space $H$ 

We recall the Pauli matrices \\

$\sigma_x = \left ( \begin{array}{l}
                0~~~~ 1 \\
                1~~~~ 0
             \end{array} \right )~~~~ \sigma_y = \left ( \begin{array}{l}
                0~~ -i \\
                i~~~~ \, 0
             \end{array} \right )~~~~  \sigma_z = \left ( \begin{array}{l}
                1~~~~ \, 0 \\
                0~~ -1
             \end{array} \right )$ \\

and the Hadamard matrix \\

$H = ( 1 / \sqrt 2 ) \left ( \begin{array}{l}
                1~~~~ 1 \\
                1~~ -1
             \end{array} \right )$ \\

which satisfy \\

$\sigma_x^2 = \sigma_y^2 = \sigma_z^2 = -i \sigma_x \sigma_y \sigma_z = H^2 = I = \left ( \begin{array}{l}
                1~~~~ 0 \\
                0~~~~ 1
             \end{array}
              \right )$ \\ \\

$ \sigma_x \sigma_y = \left ( \begin{array}{l}
                i~~~~\, 0 \\
                0~~ -i
             \end{array}
              \right ) = i \sigma_z,~~~~ \sigma_y \sigma_x = \left ( \begin{array}{l}
                -i~~ \, 0 \\
                \, \, 0~~~~ i
             \end{array}
              \right ) = -i \sigma_z$ \\ \\

$ \sigma_x \sigma_z = \left ( \begin{array}{l}
                0~~ -1 \\
                1~~~~ 0
             \end{array}
              \right ) = i \sigma_y,~~~~ \sigma_z \sigma_x = \left ( \begin{array}{l}
                \, \, 0~~~~ \, 1 \\
                -1 ~~~~ 0
             \end{array}
              \right ) = -i \sigma_y$ \\ \\

$ \sigma_y \sigma_z = \left ( \begin{array}{l}
                0~~~~ i \\
                i~~~~ 0
             \end{array}
              \right ) = i \sigma_x,~~~~ \sigma_z \sigma_y = \left ( \begin{array}{l}
                \, \, 0~~ \, -i \\
                -i~~~~ 0
             \end{array}
              \right ) = -i \sigma_x$ \\ \\

Further, we have \\

$\sigma_x ( | \, 0 > ) = | \, 1 >,~~ \sigma_x ( | \, 1 > ) = | \, 0 > \\ \\
\sigma_y ( | \, 0 > ) = i | \, 1 >,~~ \sigma_y ( | \, 1 > ) = -i | \, 0 > \\ \\
\sigma_z ( | \, 0 > ) = | \, 0 >,~~ \sigma_z ( | \, 1 > ) = - | \, 1 > \\ \\
H ( | \, 0 > ) = ( | \, 0 > +  | \, 1 > ) / \sqrt 2,~~ H ( | \, 1 > ) = ( | \, 0 > - | \, 1 >) / \sqrt 2$ \\

We take $\alpha = +$, thus the usual tensor product $\bigotimes$ is in fact $\bigotimes_{+, +}$. \\

1) Let us now take a Bell state \\

$| \, \psi > \, = ( | \, 1 > | \, 0 > + | \, 0 > | \, 1 > ) / \sqrt 2 = \\ \\
~~~~ = ( | \, 1 > \otimes | \, 0 > + | \, 0 > \otimes | \, 1 > ) / \sqrt 2 = \\ \\
~~~~ = ( | \, 1 > \otimes_{+, +} | \, 0 > + | \, 0 > \otimes_{+, +} | \, 1 > ) / \sqrt 2 \in H \bigotimes_{+, +} H = H \bigotimes H$ \\

and apply to it the mapping $\lambda_{\alpha, {\cal A}}$ in (33) for various choices of the operators $A, B \in {\cal A} = \{ \sigma_x, \sigma_y, \sigma_z, H \}$. \\

1.1) For $A = B = \sigma_x$, thus with ${\cal A} = \{ \sigma_x \}$, we have in $H \bigotimes_{+, A} H$ \\

$\lambda_{\alpha, {\cal A}} ( | \, \psi > \, ) = ( \sigma_x, \sigma_x ) ( | \, \psi > \, ) = \\ \\
= [ ( \sigma_x ( | \, 1 > ) )  ( \sigma_x ( | \, 0 > ) ) + ( \sigma_x ( | \, 0 > ) )  ( \sigma_x ( | \, 1 > ) ) ] / \sqrt 2 = \\ \\
= [ | \, 0 > ) | \, 1 > ) + | \, 1 > ) | \, 0 > ] / \sqrt 2 = | \, \psi >$ \\

1.2) For $A = B = \sigma_y$, thus with ${\cal A} = \{ \sigma_y \}$, we have in $H \bigotimes_{+, A} H$ \\

$\lambda_{\alpha, {\cal A}} ( | \, \psi > \, ) = ( \sigma_y, \sigma_y ) ( | \, \psi > \, ) = \\ \\
= [ ( \sigma_y ( | \, 1 > ) )  ( \sigma_y ( | \, 0 > ) ) + ( \sigma_y ( | \, 0 > ) )  ( \sigma_y ( | \, 1 > ) ) ] / \sqrt 2 = \\ \\
= [ ( -i | \, 0 > )  ( i | \, 1 > ) ) + ( i | \, 1 > ) ( -i | \, 0 > ) ] / \sqrt 2 = | \, \psi >$ \\

1.3) For $A = B = \sigma_z$, thus with ${\cal A} = \{ \sigma_z \}$, we have in $H \bigotimes_{+, A} H$ \\

$\lambda_{\alpha, {\cal A}} ( | \, \psi > \, ) = ( \sigma_z, \sigma_z ) ( | \, \psi > \, ) = \\ \\
= [ ( \sigma_z ( | \, 1 > ) )  ( \sigma_z ( | \, 0 > ) ) + ( \sigma_z ( | \, 0 > ) )  ( \sigma_z ( | \, 1 > ) ) ] / \sqrt 2 = \\ \\
= [ - | \, 1 > )  | \, 0 > ) - | \, 0 > ) | \, 0 > ) ] / \sqrt 2 = - | \, \psi >$ \\

1.4) For $A = \sigma_x, B = \sigma_y$, thus with ${\cal A} = \{ \sigma_x, \sigma_y \}$, we have in $H \bigotimes_{+, A} H$ \\

$\lambda_{\alpha, {\cal A}} ( | \, \psi > \, ) = ( \sigma_x, \sigma_y ) ( | \, \psi > \, ) = \\ \\
= [ ( \sigma_x ( | \, 1 > ) )  ( \sigma_y ( | \, 0 > ) ) + ( \sigma_x ( | \, 0 > ) )  ( \sigma_y ( | \, 1 > ) ) ] / \sqrt 2 = \\ \\
= [ | \, 0 > ) ( i | \, 1 > ) ) - | \, 1 > ( i | \, 0 > ) ) ] / \sqrt 2 = i | \, \chi >$ \\

where \\

$| \, \chi > \, = ( | \, 0 > ) | \, 1 > ) - | \, 1 >  | \, 0 > ) / \sqrt 2$ \\

is a Bell state. \\

1.5) For $A = \sigma_x, B = \sigma_z$, thus with ${\cal A} = \{ \sigma_x, \sigma_z \}$, we have in $H \bigotimes_{+, A} H$ \\

$\lambda_{\alpha, {\cal A}} ( | \, \psi > \, ) = ( \sigma_x, \sigma_z ) ( | \, \psi > \, ) = \\ \\
= [ ( \sigma_x ( | \, 1 > ) )  ( \sigma_z ( | \, 0 > ) ) + ( \sigma_x ( | \, 0 > ) )  ( \sigma_z ( | \, 1 > ) ) ] / \sqrt 2 = \\ \\
= [ | \, 0 > ) | \, 0 > ) - | \, 1 > | \, 1 > ) ] / \sqrt 2$ \\

which is a Bell state. \\

1.6) For $A = \sigma_y, B = \sigma_z$, thus with ${\cal A} = \{ \sigma_y, \sigma_z \}$, we have in $H \bigotimes_{+, A} H$ \\

$\lambda_{\alpha, {\cal A}} ( | \, \psi > \, ) = ( \sigma_y, \sigma_z ) ( | \, \psi > \, ) = \\ \\
= [ ( \sigma_y ( | \, 1 > ) )  ( \sigma_z ( | \, 0 > ) ) + ( \sigma_y ( | \, 0 > ) )  ( \sigma_z ( | \, 1 > ) ) ] / \sqrt 2 = \\ \\
= [ ( -i | \, 0 > ) ) | \, 0 > ) - ( i | \, 1 > | \, 1 > ) ] / \sqrt 2 = -i | \, \varphi >$ \\

where \\

$| \, \varphi > \, = | \, 0 > ) ) | \, 0 > ) + | \, 1 > | \, 1 >$ \\

is a Bell state. \\

1.7) For $A = B = H$, thus with ${\cal A} = \{ H \}$, we have in $H \bigotimes_{+, A} H$ \\

$\lambda_{\alpha, {\cal A}} ( | \, \psi > \, ) = ( H, H ) ( | \, \psi > \, ) = \\ \\
= [ ( H ( | \, 1 > ) )  ( H ( | \, 0 > ) ) + ( H ( | \, 0 > ) )  ( H ( | \, 1 > ) ) ] / \sqrt 2 = \\ \\
= [ ( ( | \, 0 > ) - | \, 1 > ) / \sqrt 2 ) ( ( | \, 0 > ) + | \, 1 > ) / \sqrt 2 ) + \\ \\
 ~~~~~~~~~~ + ( ( | \, 0 > ) + | \, 1 > ) / \sqrt 2 ) ( ( | \, 0 > ) - | \, 1 > ) / \sqrt 2 ) ] / \sqrt 2 = \\ \\
= [ ( | \, 0 > ) - | \, 1 > ) ( | \, 0 > ) + | \, 1 > ) + \\ \\
 ~~~~~~~~~~ + ( | \, 0 > ) + | \, 1 > ) ( | \, 0 > ) - | \, 1 > ) ] / ( 2 \sqrt 2 ) = \\ \\ 
 = ( | \, 0 > ) | \, 0 > ) - | \, 1 > ) | \, 1 > ) ) \sqrt 2$ \\

which is a Bell state.

\bigskip

Discussion on this paper was started during the visit of A. Khrennikov to University of Pretoria, November 2010, and completed
during the visits of E. Rosinger to Linnaeus University, June and September 2011. This research was supported by the joint project
of Swedish and South-African Research Councils, ``Non-Archimedean analysis: from fundamentals to applications.''


\begin{thebibliography}{99}

\bibitem{KC} A.   Khrennikov, {\it Ubiquitous  Quantum Structure: from
Psychology to Finances,} Springer, Berlin-Heidelberg-New York,
2010.


\bibitem{A} M. Asano, M. Ohya, Y. Togawa and A. Khrennikov, I. Basieva, 
``Quantum-like Representation of Bayesian Inference'', 
in \emph{Advances in Quantum Theory}, 
AIP Conf. Proc. 1327, American Institute of Physics, New York, 2011,  pp. 57-62.

\bibitem{R} E. E Rosinger, physics.gen-ph-0701116v8.

\bibitem{NA} A. Khrennikov, {\it Non-Archimedean analysis: quantum
paradoxes, dynamical systems and biological models,} Kluwer,
Dordreht, 1997.

\bibitem{R1} E. E Rosinger, A. Khrennikov,  ``Beyond Archimedean Space-Time Structure'',
in \emph{Advances in Quantum Theory},  AIP Conf. Proc. 1327,
 American Institute of Physics, New York, 2011,  pp. 520-526.


\end{thebibliography}
\end{document}